\documentclass[copyright,creativecommons]{eptcs}
\providecommand{\event}{T. Neary, D. Woods, A.K. Seda and N. Murphy (Eds.):\\ The Complexity of Simple Programs 2008.}
\providecommand{\volume}{1}
\providecommand{\anno}{2009}
\providecommand{\firstpage}{67}
\usepackage{breakurl}
\usepackage{breakurl}
\usepackage{amsmath}
\usepackage{amssymb}

\newcommand{\Z}{\mathbb{Z}}
\newcommand{\N}{\mathbb{N}}
\newcommand{\calT}{\mathcal{T}}
\newcommand{\setl}[2]{\left\{\ \left. #1 \ \right|\ #2 \ \right\}}
\newcommand{\setr}[2]{\left\{\ #1 \ \left|\ #2 \ \right. \right\}}
\newcommand{\kth}{{k^{\mathrm{th}}}}

\title{Limitations of Self-Assembly at Temperature One (extended abstract)}
\author{David Doty
\institute{Department of Computer Science}
\institute{Iowa State University}
\institute{Ames, IA 50011, USA}
\email{ddoty@cs.iastate.edu}
\and
Matthew J. Patitz
\institute{Department of Computer Science}
\institute{Iowa State University}
\institute{Ames, IA 50011, USA}
\email{mpatitz@cs.iastate.edu}
\and
Scott M. Summers
\institute{Department of Computer Science}
\institute{Iowa State University}
\institute{Ames, IA 50011, USA}
\email{summers@cs.iastate.edu}
}
\def\titlerunning{Limitations of Self-Assembly at Temperature One}
\def\authorrunning{D. Doty, M. Patitz and S. Summers}
\begin{document}
\maketitle

\begin{abstract}
We prove that if a set $X \subseteq \Z^2$
weakly self-assembles at temperature 1 in a deterministic (Winfree) tile
assembly system satisfying a natural condition known as \emph{pumpability}, then $X$ is a finite union of semi-doubly periodic sets. This shows that only the most simple of infinite shapes and patterns can be constructed using pumpable temperature 1 tile assembly systems, and gives evidence for the thesis that temperature 2 or higher is required to carry out general-purpose computation in a tile assembly system. 
Finally, we show that general-purpose computation \emph{is} possible at temperature 1 if negative glue strengths are allowed in the tile assembly model.
\end{abstract}

\section{Introduction}
Self-assembly is a bottom-up process by which a small number of
fundamental components automatically coalesce to form a target
structure. In 1998, Winfree \cite{Winf98} introduced the (abstract)
Tile Assembly Model (TAM) -- an ``effectivization'' of Wang tiling
\cite{Wang61,Wang63} -- as an over-simplified mathematical model of
the DNA self-assembly pioneered by Seeman \cite{Seem82}. In
the TAM, the fundamental components are un-rotatable, but
translatable square ``tile types'' whose sides are labeled with glue
``colors'' and ``strengths.'' Two tiles that are placed next to each
other \emph{interact} if the glue colors on their abutting sides
match, and they \emph{bind} if the strength on their abutting sides
matches with total strength at least a certain ambient
``temperature,'' usually taken to be 1 or 2.

Despite its deliberate over-simplification, the TAM is a
computationally and geometrically expressive model at temperature 2. The reason is that, at temperature 2, certain tiles are not
permitted to bond until \emph{two} tiles are already present to
match the glues on the bonding sides, which enables cooperation
between different tile types to control the placement of new tiles.
Winfree \cite{Winf98} proved that at temperature 2 the TAM is
computationally universal and thus can be directed algorithmically.

In contrast, we aim to prove that the lack of cooperation at
temperature 1 inhibits the sort of complex behavior possible at
temperature 2. Our main theorem concerns the \emph{weak
self-assembly} of subsets of $\Z^2$ using temperature 1 tile
assembly systems. Informally, a set $X \subseteq \Z^2$ weakly
self-assembles in a tile assembly system $\calT$ if some of the tile
types of $\calT$ are painted black, and $\calT$ is guaranteed to
self-assemble into an assembly $\alpha$ of tiles such that $X$ is
precisely the set of integer lattice points at which $\alpha$
contains black tile types. As an example, Winfree \cite{Winf98}
exhibited a temperature 2 tile assembly system, directed by a clever
XOR-like algorithm, that weakly self-assembles a well-known set, the
discrete Sierpinski triangle, onto the first quadrant. Note that the
underlying \emph{shape} (set of all points containing a tile,
whether black or not) of Winfree's construction is an infinite
canvas that covers the entire first quadrant, onto which a more
sophisticated set, the discrete Sierpinski triangle, is painted.

We show that under a plausible assumption, temperature 1 tile
systems weakly self-assemble only a limited class of sets. To prove
our main result, we require the hypothesis that the tile system is
\emph{pumpable}. Informally, this means that every sufficiently long
path of tiles in an assembly of this system contains a segment in
which the same tile type repeats (a condition clearly implied by the
pigeonhole principle), and that furthermore, the subpath between
these two occurrences can be repeated indefinitely (``pumped'')
along the same direction as the first occurrence of the segment,
without ``colliding'' with a previous portion of the path. We give a
counterexample of a path in which the same tile type appears
twice, yet the segment between the appearances cannot be pumped
without eventually resulting in a collision that prevents additional
pumping. The hypothesis of pumpability states (roughly) that in
every sufficiently long path, despite the presence of some repeating
tiles that cannot be pumped, \emph{there exists} a segment in which
the same tile type repeats that \emph{can} be pumped. In the
above-mentioned counterexample, the paths constructed to create a
blocked segment always contain some previous segment that \emph{is}
pumpable. We conjecture that this phenomenon, pumpability, occurs in
every temperature 1 tile assembly system that produces a unique infinite structure.

A \emph{semi-doubly periodic} set $X \subseteq \Z^2$ is a set of
integer lattice points with the property that there are three
vectors $\vec{b}$ (the ``base point'' of the set), $\vec{u}$, and
$\vec{v}$ (the two periods of the set), such that $X = \setl{\vec{b}
+ n \cdot \vec{u} + m \cdot \vec{v}}{n,m\in\N}$. That is, a semi-doubly periodic set is a set that repeats infinitely
along two vectors (linearly independent vectors in the
non-degenerate case), starting at some base point $\vec{b}$. We show
that any directed, pumpable, temperature 1 tile assembly system
weakly self-assembles a set $X \subseteq \Z^2$ that is a finite
union of semi-doubly periodic sets.

It is our contention that weak self-assembly captures the intuitive
notion of what it means to ``compute'' with a tile assembly system.
For example, the use of tile assembly systems to build shapes is
captured by requiring all tile types to be black, in order to ask
what set of integer lattice points contain any tile at all
(so-called \emph{strict self-assembly}). However, weak self-assembly is a
more general notion. For example, Winfree's above mentioned result
shows that the discrete Sierpinski triangle weakly self-assembles at
temperature 2 \cite{RoPaWi04}, yet this shape does not strictly
self-assemble at \emph{any} temperature \cite{jSSADST}. Hence weak
self-assembly allows for a more relaxed notion of set building, in
which intermediate space can be used for computation, without
requiring that the space filled to carry out the computation also
represent the final result of the computation.

As another example, there is a standard construction \cite{Winf98}
by which a single-tape Turing machine may be simulated by a
temperature 2 tile assembly system. Regardless of the semantics of
the Turing machine (whether it decides a language, enumerates a
language, computes a function, etc.), it is routine to represent the
result of the computation by the weak self-assembly of some set. For
example, Patitz and Summers \cite{SADS} showed that for any
decidable language $A \subseteq \N$, $A$'s projection along the
$X$-axis $\left( \textmd{the set}\setl{(x,0) \in \N^2}{x \in A}\right)$ weakly
self-assembles in a temperature 2 tile assembly system.
As another
example, if a Turing machine computes a function $f:\N\to\N$, it is
routine to design a tile assembly system based on Winfree's
construction such that, if the seed assembly is used to encode the
binary representation of a number $n\in\N$, then the tile assembly
system weakly self-assembles the set $$\setr{(k,0) \in
\N^2}{\begin{split}\text{the $\kth$ least significant bit of the}\\ \text{binary representation of $f(n)$ is 1}\end{split}}.$$

Our result is motivated by the thesis that if a tile assembly system
can reasonably be said to ``compute'', then the result of this
computation can be represented in a straightforward manner as a set
$X \subseteq \Z^2$ that weak self-assembles in the tile assembly
system, or a closely related tile assembly system. Our examples
above provide evidence for this thesis, although it is as informal
and unprovable as the Church-Turing thesis.  On the basis of this
claim, and the triviality of semi-doubly periodic sets, we
conclude that our main result implies that directed, pumpable,
temperature 1 tile assembly systems are incapable of general-purpose
deterministic computation, without further relaxing the model.

\subsubsection*{Acknowledgments}
We wish to thank Maria Axenovich, Matt Cook, and Jack Lutz for
useful discussions, and anonymous referees for corrections. We would especially like to thank Niall Murphy,
Turlough Neary, Anthony K. Seda and Damien Woods for inviting us to present a
preliminary version of this research at the International Workshop
on The Complexity of Simple Programs, University College Cork,
Ireland on December 6th and 7th, 2008.

\bibliographystyle{eptcs}


\end{document}